\documentstyle[12pt,psfig,epsf]{article}
\newcommand{\be}{\begin{equation}}
\newcommand{\ee}{\end{equation}}

\newcommand{\beqa}{\begin{eqnarray}}
\newcommand{\eeqa}{\end{eqnarray}}

\newcommand{\eqref}[1]{(\ref{#1})}

\def\boxit#1{\vbox{\hrule\hbox{\vrule\kern8pt
\vbox{\hbox{\kern8pt}\hbox{\vbox{#1}}\hbox{\kern8pt}}
\kern8pt\vrule}\hrule}}
\def\mathboxit#1{\vbox{\hrule\hbox{\vrule\kern8pt\vbox{\kern8pt
\hbox{$\displaystyle #1$}\kern8pt}\kern8pt\vrule}\hrule}}

\def\IB{\relax\hbox{$\inbar\kern-.3em{\rm B}$}}
\def\IC{\relax\hbox{$\inbar\kern-.3em{\rm C}$}}
\def\ID{\relax\hbox{$\inbar\kern-.3em{\rm D}$}}
\def\IE{\relax\hbox{$\inbar\kern-.3em{\rm E}$}}
\def\IF{\relax\hbox{$\inbar\kern-.3em{\rm F}$}}
\def\IG{\relax\hbox{$\inbar\kern-.3em{\rm G}$}}
\def\IGa{\relax\hbox{${\rm I}\kern-.18em\Gamma$}}
\def\IH{\relax{\rm I\kern-.18em H}}
\def\IK{\relax{\rm I\kern-.18em K}}
\def\IL{\relax{\rm I\kern-.18em L}}
\def\IP{\relax{\rm I\kern-.18em P}}
\def\IR{\relax{\rm I\kern-.18em R}}
\def\IZ{\relax\ifmmode\mathchoice
{\hbox{\cmss Z\kern-.4em Z}}{\hbox{\cmss Z\kern-.4em Z}}
{\lower.9pt\hbox{\cmsss Z\kern-.4em Z}} {\lower1.2pt\hbox{\cmsss
Z\kern-.4em Z}}\else{\cmss Z\kern-.4em Z}\fi}

\def\II{\relax{\rm I\kern-.18em I}}

\begin{document}

\hfill CERN-PH-TH/2004-026

\hfill NRCPS-HE-2004-01

\vspace{5cm}
\begin{center}
{\LARGE Physical Fock Space of Tensionless Strings

}

\vspace{2cm}
{\sl Ignatios Antoniadis$^{1,a,}$${}^\dagger$\footnote{${}^\dagger$ On leave of absence
from CPHT {\'E}cole Polytechnique, F-91128, Palaiseau Cedex,France.} and  George Savvidy$^{1,b,}{}^{\dagger\dagger}$
\footnote{${}^{\dagger\dagger}$On leave of absence
from Demokritos National Research Center$^{2}$.}

\bigskip
\centerline{${}^1$ \sl Department of Physics, CERN Theory Division CH-1211 Geneva 23, Switzerland}
\bigskip
\centerline{${}^2$ \sl Demokritos National Research Center, Ag. Paraskevi, GR-15310, Athens}
\bigskip
\centerline{$^a$~{\footnotesize\it ignatios.antoniadis@cern.ch},
$^b$~{\footnotesize\it georgios.savvidis@cern.ch}}
\bigskip
}
\end{center}
\vspace{60pt}

\centerline{{\bf Abstract}}

\vspace{12pt}

\noindent
We study the physical Fock space of the tensionless string theory
with perimeter action which has pure massless spectrum.
The states are classified by the Wigner's little group
for massless particles. The ground state contains infinite many massless fields
of fixed helicity, the excitation levels realize CSR representations.
We demonstrate that the first and the second excitation levels are physical null states.


\newpage

\pagestyle{plain}

\section{{\it Introduction}}

A string model which is based on the concept of surface
perimeter was suggested in  \cite{Savvidy:dv,Savvidy:2003fx,geo}.
At the classical level the model is {\it tensionless},
because for the flat Wilson loop the action is equal to its perimeter \cite{geo}.
In the recent articles one of the authors has found that it has
pure massless spectrum of infinitely many integer spin fields \cite{Savvidy:dv,Savvidy:2003fx}
\footnote{This model
differs from the tensionless string models based on Schild's work
on "null" string \cite{Schild:vq,deVega:1987hu,Gasperini:1991rv,Lousto:1996hg}
with its most likely continuous spectrum
\cite{Lindstrom:1990qb,
DeVega:1992tm,Lizzi:1994rn}.}.

The solution of the corresponding two-dimensional world-sheet CFT for the
canonically conjugate operators
$X^{\mu} = (X^{\mu}_L + X^{\mu}_R )/2$ and
$\Pi^{\mu} = (\Pi^{\mu}_L + \Pi^{\mu}_R )/2$ is \cite{Savvidy:dv}:
\beqa\label{solution}
X^{\mu}_{L} = x^{\mu} +
{1\over m}\pi^{\mu}\zeta^{+} +
i\sum_{n \neq 0}  {1\over n }~ \beta^{\mu}_{n} e^{-in\zeta^{+}},\nonumber\\
\Pi^{\mu}_{L}=   m e^{\mu} +  k^{\mu}\zeta^{+} + i\sum_{n \neq 0}
{1\over n }~ \alpha^{\mu}_{n} e^{-in\zeta^{+}},
\eeqa
where~$k^{\mu}$ is momentum operator,
$\alpha_n$,$\beta_n$ are oscillators satisfying the
following commutator relations $[x^{\mu},
k^{\nu}] = i\eta^{\mu\nu},~[\alpha^{\mu}_{n},
\beta^{\nu}_{l}]=n~\eta^{\mu\nu}\delta_{n+l,0}$
and $\zeta^{\pm} = \tau \pm \sigma$.
A similar expansion also holds for the right
moving modes $X^{\mu}_{R},~ \Pi^{\mu}_{R}$ with the
oscillators $\tilde{\beta}_n$,~$\tilde{\alpha}_n$.
The essential new feature of this theory is the appearance of
the additional zero modes $[e^{\mu},\pi^{\nu}]=i\eta^{\mu\nu}$, which
means that the space-time wave function is  a function of two continuous
variables $k^{\mu}$ and $e^{\mu}$:
$$
\Psi_{Phys}= \Psi(k,e).
$$
It was suggested in \cite{Savvidy:dv} that $e^{\mu}$ should be
interpreted as a polarization vector. Its conjugate operator
is  defined as: $\pi^{\mu} = i \partial/ \partial e_{\mu}$.

Our aim is to study spectral properties of the
model on higher levels of physical Fock space
in analogy with the standard string theory
\cite{Brower:1971qr,DelGiudice:1971fp,Brower:ev,
Brower:1972wj,Goddard:iy,Goddard:1972ky}.
This is quite possible using the mode expansion of the
world-sheet fields and exploring corresponding new gauge symmetries.
The tensionless string theory is invariant with respect to the
standard conformal group with generators
$$
L_n =\sum_{l}:\alpha_{n-l} \cdot\beta_{l}: ~~~~~~n=0,\pm 1 , \pm 2
$$
and with respect to the infinite-dimensional
Abelian transformations generated by the operator
$\Theta = \Pi^{2}_{\mu} -1$ with the following components:
$$
k^2,~~~ k \cdot e,~~~ k \cdot \alpha_n,~~~ k\cdot
\tilde{\alpha_n},~~~ \Theta_{nl} ~~~~~~n,l=0,\pm 1 , \pm 2...,
$$
corresponding to the coefficients of $\tau^2$, $\tau$, $\tau$ with
exponentials and Fourier modes, respectively.
Therefore in covariant quantization scheme
the space-time equations, which define physical Fock space, have
the following form \cite{Savvidy:dv,Savvidy:2003fx}:
\be\label{first}
\left( \begin{array}{l}
  k^2\\
  k\cdot e\\
  k\cdot \alpha_n\\
  k\cdot \tilde{\alpha_n}\\
  \Theta_{nl}\\
  L_n   \\
  \tilde{L_n}
\end{array} \right)\Psi_{Phys}=0,~~~~~~~n,l = 0,1,2,....
\ee
As one can clearly see from the first equation in (\ref{first}),
the spectrum of the model is purely massless: $k^2 =0$.

The zero mode conformal operator is given by the expression
\footnote{We shall use physical units in which $m=1$.}
$$
L_0 = (k \cdot \pi) ~+ ~\hat{\Xi},
$$
where $\hat{\Xi} =  \sum_{n \neq 0}~ \alpha_{-n} \beta_n$
is the number operator with eigenvalues $\Xi = 0,1,2...$. This
operator allows to represent the whole Fock space ${{\cal F}}$ as a sum
of relativistically invariant subspaces
$$
{{\cal F}} = \sum^{\infty}_{\Xi=0} \oplus{{\cal F}}_{\Xi}
$$
and to study each subspace ${{\cal F}}_{\Xi}$ separately.
A natural question that arises is whether
the physical Hilbert space
on each subspace ${{\cal F}}_{\Xi}$ is positive-definite,
i.e. ghost-free in analogy with the well known No-ghost theorem
\cite{Brower:1971qr,DelGiudice:1971fp,Brower:ev,
Brower:1972wj,Goddard:iy,Goddard:1972ky}
in the standard string theory.

As it was demonstrated in \cite{Savvidy:2003fx}, the first
two levels $\Xi =0,1$ corresponding to the ground state and the first excitation
state, are well defined and have no negative norm
waves. The vacuum state ${{\cal F}}_{0}$ is infinitely degenerate
and contains massless particles of increasing tensor structure
$A^{\mu_1 ,..., \mu_s}(k)$ for $s=1,2,...$, while the first
level ${{\cal F}}_{1}$ happens to be a physical null state.

We shall demonstrate below that the {\it second level} wave function,
$\Xi =2$, also represents a physical null state, therefore one can guess
that all excitation levels
$\Xi =1,2,3,...$ are {\it physical null states}. The general proof
of this conjecture is still to be found despite the fact that some
pattern already appears in the first two excitation levels, which explains
why all excitations are physical zero norm states. The main
reason is, that the polarization tensors must fulfill two constraints
which follow from the main space-time equations (\ref{first}):
i) they have to be transverse $k_{\mu_1}\xi_{\mu_1 ,...\mu_s} =0$
and at the same time ii) to be longitudinal $e_{\mu_1}\xi_{\mu_1 ,...\mu_s} =0$.
This leaves no options and the only solution has the form:
$$\xi_{\mu_1 ,...\mu_s} =
\xi(k,e)~k_{\mu_1}...k_{\mu_s}
$$ and has zero norm!

In the next section we shall review the solution of
the basic space-time equations (\ref{first}) in Gupta-Bleuler
quantization scheme for
the ground state and the first excitation states \cite{Savvidy:dv,Savvidy:2003fx}.
Then we shall turn to the study of the second level wave function.

\section{\it $\Theta$-Algebra }

The symmetries of the model are governed by
the conformal operators $L_n$ and the new operators $\Theta_{n,l}$:
\be
L_{n}  = <e^{in\zeta^+} :P^{\mu}_{L}~\partial_{+}X^{\mu}_{L}: >,~~~~
\Theta_{n,l}  =  <e^{in\zeta^+ + il\zeta^-} :  \Pi^{\mu}~\Pi^{\mu} - 1  :>
\label{constcompo}
\ee
where $P^{\mu} = \partial_{\tau} \Pi^{\mu} = (P^{\mu}_{L} +P^{\mu}_{R})/2$
is the momentum operator defined by the mode
expansion (\ref{solution});~ thus
\beqa\label{fullalgebra}
L_{n} &=&\sum_{l}:\alpha_{n-l} \cdot\beta_{l}:~~~~~
\tilde{L}_{n} =\sum_{l}: \tilde{\alpha}_{n-l} \cdot \tilde{\beta}_{l}:\nonumber\\
\Theta_{0,0} &=& ( e^{2} -1)
+ \sum_{n \neq 0}
{1\over 4 n^2}:(\alpha_{-n}~\alpha_{n} +
\tilde{\alpha}_{-n}~\tilde{\alpha}_{n}):\nonumber\\
\Theta_{n,0} &=& {i\over n}~e \cdot \alpha_{n}
-{1\over 4}\sum_{l \neq 0,n}
{1\over (n-l)l}:\alpha_{n-l}\cdot\alpha_{l}:~~~~~~~~~~~n=\pm1,\pm2,..\nonumber\\
\Theta_{0,n} &=&{i\over n}~e \cdot \tilde{\alpha}_{n}
-{1\over 4}\sum_{l \neq 0,n}
{1\over (n-l)l}:\tilde{\alpha}_{n-l}\cdot\tilde{\alpha}_{l}:
~~~~~~~~~~~n=\pm1,\pm2,..\nonumber\\
\Theta_{n,l} &=& -{1\over 2 n l}:\alpha_{n}\cdot\tilde{\alpha}_{l}:
~~~~~~~~~~~~~~~~~~~~~~~~~~~n,l= \pm 1, \pm 2,....
\eeqa
The conformal algebra has here its classical form, but
with twice larger central charge
\be\label{oldalgebra}
[L_n ~,~ L_l] = (n-l) L_{n+l} + {D\over 6}(n^3 -n)\delta_{n+l,0}
\ee
and with a similar expression for the right movers $\tilde{L}_n$.
The reason that the central charge is twice bigger than in the
standard bosonic string theory $2 \times {D\over 12} = {D\over 6}$
is simply because we have two left and
two right moving fields
\footnote{Such doubling of modes is reminiscent of the bosonic part of the
${\cal N}=2$ superstring \cite{Ademollo:1974fc}.
Here the coordinate field $X$ has simply
two sets of commuting oscillators and the conjugate oscillators are
described by a separate field $\Pi_{\mu}$.}.

The full extended gauge symmetry algebra of constraints
(\ref{constcompo}) takes the form
\beqa\label{newalgebra1}
~[~L_n  , \Theta_{0,0}] &=& -2n  \Theta_{n,0}~~~~~~~~~~~~~~~~~~~~~~~~~~~~
[\tilde{L}_n  , \Theta_{0,0}] = -2n \Theta_{0,n} \nonumber\\
~[~L_n , \Theta_{l,0}] &=& -(n+l) \Theta_{n+l,0}~~~~~~~~~~~~~~~~~~~~
[\tilde{L}_n , \Theta_{l,0}] = -2n \Theta_{l,n}\nonumber\\
~[~L_n , \Theta_{0,l}] &=& -2n\Theta_{n,l}~~~~~~~~~~~~~~~~~~~~~~~~~~~~
[\tilde{L}_n ,  \Theta_{0,l}]
= -(n+l) \Theta_{0,n+l} \nonumber\\
~[~L_{n} , \Theta_{m,l}] &=& -(n+m) \Theta_{n+m,l}~~~~~~~~~~~~~~~~~~
[\tilde{L}_n , \Theta_{m,l}]
= -(n+l) \Theta_{m,n+l} \nonumber\\
~[~L_{n} , ~k^2 ~] &=& 0~~~~~~~~~~~~~~~~~~~~~~~~~~~~~~~~~~~~~~
[~\tilde{L}_n ,~ k^2~]
= 0 \nonumber\\
~[~L_{n} , k\cdot e] &=& - i~k\cdot \alpha_n~~~~~~~~~~~~~~~~~~~~~~~~~~~~
[~\tilde{L}_n , k\cdot e]
= - i ~k\cdot \tilde{\alpha}_n \nonumber\\
~[L_{n} , k\cdot \alpha_l] &=& -l~~k\cdot \alpha_{n+l}~~~~~~~~~~~~~~~~~~~~~~~~~
[\tilde{L}_n , k\cdot \tilde{\alpha}_l]
= -l~ ~k\cdot \tilde{\alpha}_{n+l} ,
\eeqa
where we have included commutators with the operators ~$~k^2 , ~~k\cdot e ,~~
k\cdot \alpha_l ,~~ k\cdot \tilde{\alpha}_l$, because they are constituents
of the $\tau$ dependent part of the operator $\Theta = \Pi^2 -1$, as one
can see substituting the solution (\ref{solution}) into the definition
of the operator $\Pi_{\mu} = (\Pi^{\mu}_L + \Pi^{\mu}_R)/2$.
One should stress that it is an essentially Abelian extension because
\be\label{newalgebra}
 [\Theta_{n,m} , \Theta_{l,p}] =
0,~~~~~n,m,l,p=0, \pm 1,\pm 2,...
\ee
The relations (\ref{oldalgebra}), (\ref{newalgebra1}) and (\ref{newalgebra})
define an Abelian extension of the conformal algebra and the equations
(\ref{fullalgebra}) realize its oscillator representation.

\subsection{\it Physical Fock space}
The spectrum of this theory is pure massless and the subspaces
${{\cal F}}_{N}$ defined by the operator
$L_0 =(k \cdot \pi) + \hat{N}$  do
not correspond to different mass levels, as it was in the standard
string theory, where $L_0 = k^2 + \hat{N}$,
but classify the states with respect to the eigenvalues
of the operator $(k \cdot \pi)$ \cite{Savvidy:2003fx}.
This operator is equal to the length of the
highest Casimir operator $W= (k \cdot \pi)^2$ of the Poincar\'e group
and defines fixed helicity states, when $W=0$ and
continuous spin representations-CSR, when $W\neq 0$ \cite{wigner,brink}.

\subsection{\it Ground state,~ $\Xi =0$}
Let us first describe the ground state, ~{\bf $\Xi =0$}.
The wave function $\Psi_{0}(k,e)~ \equiv~ |k,e,0>$~
is defined by $\alpha_n ~|0,k,e>~ =$ $ ~\beta_n|0,k,e>~=~0,$ $~~~n=1,2,.. $
and the system (\ref{first})
reduces to the following four equations \cite{Savvidy:dv}:
\be\label{second}
k^2 ~\Psi_{0} = 0,~~~~e \cdot k ~\Psi_{0}=0,~~~~
(e^2 - 1)~\Psi_{0}=0,~~~~(k \cdot \pi ) ~\Psi_{0} =0 .
\ee
It follows from the third equation
that ~$\Psi_0$ ~is a function defined on a unit sphere
$e^2 =1$ with signature ~$\eta^{\mu\nu}=(-+...+)$ and can be
expanded in the corresponding basis:
$$
\Psi_{0} ~= ~|0,k,e> ~= ~A(k) ~+~ A^{\mu_1}(k)~ e_{\mu_1} ~+~
A^{\mu_1 , \mu_2}(k)~ e_{\mu_1} e_{\mu_2} +...|0,k>.
$$
The fields $A^{\mu_1 ,..., \mu_s}(k)$ describe massless
particles of increasing tensor structure and the
vacuum state in infinitely degenerate. They are symmetric
traceless tensor fields because they are harmonic functions
on a sphere $e^2 =1$, and it follows from the
last equation in (\ref{second}), that
$
k_{\mu_1}~A^{\mu_1,...\mu_s}(k) =0.
$
Thus the fields $A^{\mu_1,...\mu_s}$ are: 1)
symmetric traceless tensors of increasing rank $s=0,1,2,...$,
2) divergent free $k_{\mu_1} ~A^{\mu_1,...\mu_s} =0$
and 3) satisfying massless wave equations
$k^2 ~A^{\mu_1,...\mu_s} =0$.~
All the above conditions are sufficient to
describe integer spin fields
\cite{pauli,singh,fronsdal,Ferrara:1998jm,Haggi-Mani:2000ru,Sundborg:2000wp,
Witten:2000,Bengtsson:1983pg,Bengtsson:1983pd,Sezgin:2001zs,Dhar:2003fi,
Mikhailov:2002bp,Francia:2002aa}.
Thus the ground state is infinitely degenerate and contains
gauge particles of arbitrary large {\it fixed helicity}, $W= (k \cdot \pi )^2=0$.
Every tensor gauge field $A^{\mu_1,...\mu_s},~s=0,1,2,..$ appears
only once in the spectrum.

\subsection{\it The First Excitation states,~ $\Xi =1$}
These states correspond to the continuous spin representations
of massless little group, because for them $W = (k\cdot \pi)^2 =1$.
The wave function of the first excited state
depends on four polarization tensors.
The restriction on these tensors which follows from the space-time
equations (\ref{first}) gives the solution in the form \cite{Savvidy:2003fx}
\be
\Psi_1 =  [~\xi_{\mu\nu} \alpha^{\mu}_{-1}
\tilde{\alpha}^{\nu}_{-1} + \omega_{\mu\nu} (\alpha^{\mu}_{-1}
\tilde{\beta}^{\nu}_{-1} - \beta^{\mu}_{-1}
\tilde{\alpha}^{\nu}_{-1}) ~] |0,k,e>,
\ee
where~$k^\mu \omega_{\mu\nu} = \omega_{\mu\nu} k^\nu =0,~
e^\mu \omega_{\mu\nu} = \omega_{\mu\nu} e^\nu =0$. Assuming that the second
equation is valid for any vector $e^{\mu}$ the
unique solution of the last equations is:
$\omega^{\mu\nu} =\omega(k,e)~k_{\mu}k_{\nu}$. The tensor
$\xi_{\mu\nu}(k,e)$ remains arbitrary.
The norm of this state is~
$<\Psi_1 | \Psi_1> =-\omega^{*}_{\mu\nu}\omega^{\mu\nu} =
-\vert \omega \vert^2 ~(k^{2})^{2}
$
and therefore is equal to zero!
It is also normal to the ground state $<\Psi_0 |\Psi_1> =0$. Thus
the first excitation is a physical null state.
This consideration allows to make a conjecture that {\it all
excitations are physical null states} and therefore define
gauge parameters of the large gauge group.

\section{\it The Second Excitation State,~$\Xi =2$}
Let us consider the next level. The $L_0 |\Psi_2> =0$ equation
will take the form
\be
(k\cdot \pi)|\Psi_2>  = - 2~|\Psi_2>.
\ee
The $\Xi =2$ level contains twenty five tensors because
we have now five relevant operators in the left sector
$$
\alpha^{\mu}_{-1}  \alpha^{\nu}_{-1},~~~ \alpha^{\mu}_{-1}\beta^{\nu}_{-1},~~~
\beta^{\mu}_{-1}\beta^{\nu}_{-1},~~~\alpha^{\mu}_{-2},~~~\beta^{\mu}_{-2}
$$
and the same amount in the right sector. They form three "families" of operators
with tensor coefficients $\chi,\eta,\xi$
\beqa
|\Psi_2> ~~~~~~~~~~~~~~~~~~= ~~~~~~~~~~~~~
\chi^{(1)}_{\mu\nu} \alpha^{\mu}_{-2}  \tilde{\alpha}^{\nu}_{-2}
+ \chi^{(2)}_{\mu\nu}  \alpha^{\mu}_{-2} \tilde{\beta}^{\nu}_{-2} +
\chi^{(3)}_{\mu\nu}  \beta^{\mu}_{-2}  \tilde{\alpha}^{\nu}_{-2} +
\chi^{(4)}_{\mu\nu}  \beta^{\mu}_{-2}  \tilde{\beta}^{\nu}_{-2}+\nonumber\\
~\eta^{(1)}_{\mu\nu\lambda} \alpha^{\mu}_{-2}  \tilde{\alpha}^{\nu}_{-1} \tilde{\alpha}^{\lambda}_{-1}
+ \eta^{(2)}_{\mu\nu\lambda} \alpha^{\mu}_{-2}  \tilde{\alpha}^{\nu}_{-1}  \tilde{\beta}^{\lambda}_{-1} +
~
\eta^{(4)}_{\mu\nu\lambda}   \alpha^{\mu}_{-2}  \tilde{\beta}^{\nu}_{-1}  \tilde{\beta}^{\lambda}_{-1} +\nonumber\\
~\eta^{(5)}_{\mu\nu\lambda} \beta^{\mu}_{-2}  \tilde{\alpha}^{\nu}_{-1} \tilde{\alpha}^{\lambda}_{-1}
+ \eta^{(6)}_{\mu\nu\lambda} \beta^{\mu}_{-2}  \tilde{\alpha}^{\nu}_{-1}  \tilde{\beta}^{\lambda}_{-1} +
~
\eta^{(8)}_{\mu\nu\lambda}   \beta^{\mu}_{-2}  \tilde{\beta}^{\nu}_{-1}  \tilde{\beta}^{\lambda}_{-1} +\nonumber\\
~\eta^{(9)}_{\mu\nu\lambda} \alpha^{\mu}_{-1}  \alpha^{\nu}_{-1}  \tilde{\alpha}^{\lambda}_{-2}
+ \eta^{(10)}_{\mu\nu\lambda} \alpha^{\mu}_{-1}  \beta^{\nu}_{-1} \tilde{\alpha}^{\lambda}_{-2} +
~
\eta^{(12)}_{\mu\nu\lambda}   \beta^{\mu}_{-1} \beta^{\nu}_{-1}    \tilde{\alpha}^{\lambda}_{-2} +\nonumber\\
~\eta^{(13)}_{\mu\nu\lambda} \alpha^{\mu}_{-1}  \alpha^{\nu}_{-1}  \tilde{\beta}^{\lambda}_{-2}
+ \eta^{(14)}_{\mu\nu\lambda} \alpha^{\mu}_{-1}  \beta^{\nu}_{-1} \tilde{\beta}^{\lambda}_{-2} +
~
\eta^{(16)}_{\mu\nu\lambda}   \beta^{\mu}_{-1} \beta^{\nu}_{-1}    \tilde{\beta}^{\lambda}_{-2} +\nonumber\\
~\xi^{(1)}_{\mu\nu\lambda\rho} \alpha^{\mu}_{-1}  \alpha^{\nu}_{-1}\tilde{\alpha}^{\lambda}_{-1}\tilde{\alpha}^{\rho}_{-1}
+ \xi^{(2)}_{\mu\nu\lambda\rho} \alpha^{\mu}_{-1}  \alpha^{\nu}_{-1}\tilde{\alpha}^{\lambda}_{-1}\tilde{\beta}^{\rho}_{-1} +
~
\xi^{(4)}_{\mu\nu\lambda\rho}  \alpha^{\mu}_{-1}  \alpha^{\nu}_{-1} \tilde{\beta}^{\lambda}_{-1}\tilde{\beta}^{\rho}_{-1} +\nonumber\\
~\xi^{(5)}_{\mu\nu\lambda\rho} \alpha^{\mu}_{-1}  \beta^{\nu}_{-1}\tilde{\alpha}^{\lambda}_{-1}\tilde{\alpha}^{\rho}_{-1}
+ \xi^{(6)}_{\mu\nu\lambda\rho} \alpha^{\mu}_{-1}  \beta^{\nu}_{-1}\tilde{\alpha}^{\lambda}_{-1}\tilde{\beta}^{\rho}_{-1} +
~
\xi^{(8)}_{\mu\nu\lambda\rho}  \alpha^{\mu}_{-1}  \beta^{\nu}_{-1} \tilde{\beta}^{\lambda}_{-1}\tilde{\beta}^{\rho}_{-1} +\nonumber\\
~
~
~
~
~\xi^{(13)}_{\mu\nu\lambda\rho} \beta^{\mu}_{-1}  \beta^{\nu}_{-1}\tilde{\alpha}^{\lambda}_{-1}\tilde{\alpha}^{\rho}_{-1}
+ \xi^{(14)}_{\mu\nu\lambda\rho} \beta^{\mu}_{-1}  \beta^{\nu}_{-1}\tilde{\alpha}^{\lambda}_{-1}\tilde{\beta}^{\rho}_{-1} +
~
\xi^{(16)}_{\mu\nu\lambda\rho}  \beta^{\mu}_{-1}  \beta^{\nu}_{-1} \tilde{\beta}^{\lambda}_{-1}\tilde{\beta}^{\rho}_{-1} +\nonumber\\
|k,e,0>,\nonumber
\eeqa
where we have taken into account that the operators involved commute
$[\alpha^{\mu}_{-1},\alpha^{\nu}_{-1}]= [\beta^{\mu}_{-1},\beta^{\nu}_{-1}]=
[\alpha^{\mu}_{-1},\beta^{\nu}_{-1}]=0$.
The corresponding tensors $\chi,\eta,\xi$ are subject to the constraints
\beqa
\left( \begin{array}{l}
  k^2\\
  k \cdot e\\
  k\cdot \alpha_1\\
  k\cdot \tilde{\alpha_1}\\
  k\cdot \alpha_2\\
  k\cdot \tilde{\alpha_2}\\
  \Theta_{00}\\
  \Theta_{10}\\
  \Theta_{20}\\
  \Theta_{01}\\
  \Theta_{02}\\
  \Theta_{12}\\
  \Theta_{21}\\
  \Theta_{11}\\
  \Theta_{22}\\
    L_1 \\
  \tilde{L}_1\\
   L_2 \\
  \tilde{L}_2
\end{array} \right)|\Psi_2>=
\left( \begin{array}{l}
  k^2\\
  k\cdot e\\
  k\cdot \alpha_1\\
  k\cdot \tilde{\alpha_1}\\
  k\cdot \alpha_2\\
  k\cdot \tilde{\alpha_2}\\
  (e^2 -1) + {1\over 2} (\alpha_{-1} \alpha_1
  + \tilde{\alpha}_{-1} \tilde{\alpha}_1) + {1\over 8} (\alpha_{-2} \alpha_2
  + \tilde{\alpha}_{-2} \tilde{\alpha}_2)\\
  i ~e \cdot \alpha_1 + {1\over 8} \alpha_{-1} \alpha_2\\
  {i \over 2} ~e \cdot \alpha_2 - {1\over 4} \alpha_{1} \alpha_1\\
  i ~e \cdot \tilde{\alpha}_1 + {1\over 8} \tilde{\alpha}_{-1} \tilde{\alpha}_2\\
  {i \over 2} ~e \cdot \tilde{\alpha}_2 -
  {1\over 4} \tilde{\alpha}_{1} \tilde{\alpha}_1\\
  -{1\over 4} \alpha_1 \tilde{\alpha}_2\\
  -{1\over 4} \alpha_2 \tilde{\alpha}_1\\
    -{1\over 2} \alpha_1 \tilde{\alpha}_1\\
    -{1\over 8} \alpha_2 \tilde{\alpha}_2\\
  k \beta_1 + \pi \alpha_1 +\alpha_{-1} \beta_2 +
  \beta_{-1} \alpha_2\\
  k \tilde{\beta}_1 + \pi \tilde{\alpha}_1 + \tilde{\alpha}_{-1}
  \tilde{\beta}_2 +   \tilde{\beta}_{-1} \tilde{\alpha}_2\\
  k \beta_2  + \pi \alpha_2  +  \alpha_1 \beta_1\\
  k \tilde{\beta}_2 + \pi \tilde{\alpha}_2 + \tilde{\alpha}_{1}
  \tilde{\beta}_1
\end{array} \right)|\Psi_2>=0.
\eeqa
It is a rather complicated system of space-time equations and our aim is
to solve it. We shall start
analyzing the equation $\Theta_{00}|\Psi_1>=0$, because it is the most
restrictive on the wave function
$$\Theta_{00}|\Psi_1>=
{1\over 4} \chi^{(2)}_{\mu\nu}\alpha^{\mu}_{-2} \tilde{\alpha}^{\nu}_{-2} +
{1\over 4} \chi^{(3)}_{\mu\nu}\alpha^{\mu}_{-2} \tilde{\alpha}^{\nu}_{-2} +
{1\over 4} \chi^{(4)}_{\mu\nu}\alpha^{\mu}_{-2} \tilde{\beta}^{\nu}_{-2} +
{1\over 4} \chi^{(4)}_{\mu\nu}\beta^{\mu}_{-2} \tilde{\alpha}^{\nu}_{-2} +
$$
plus the part which is cubic in the operators
\beqa
{1\over 2} \eta^{(2)}_{\mu\nu\lambda} \alpha^{\mu}_{-2}  \tilde{\alpha}^{\nu}_{-1}  \tilde{\alpha}^{\lambda}_{-1} +
~
{1\over 2} \eta^{(4)}_{\mu\nu\lambda}   \alpha^{\mu}_{-2}  \tilde{\alpha}^{\nu}_{-1}  \tilde{\beta}^{\lambda}_{-1} +
{1\over 2} \eta^{(4)}_{\mu\nu\lambda}   \alpha^{\mu}_{-2}  \tilde{\beta}^{\nu}_{-1}  \tilde{\alpha}^{\lambda}_{-1} +
\nonumber\\
{1\over 4}\eta^{(5)}_{\mu\nu\lambda} \alpha^{\mu}_{-2}  \tilde{\alpha}^{\nu}_{-1} \tilde{\alpha}^{\lambda}_{-1}+
{1\over 4}\eta^{(6)}_{\mu\nu\lambda} \alpha^{\mu}_{-2}  \tilde{\alpha}^{\nu}_{-1}  \tilde{\beta}^{\lambda}_{-1} +
~
{1\over 4}\eta^{(8)}_{\mu\nu\lambda}   \alpha^{\mu}_{-2}  \tilde{\beta}^{\nu}_{-1}  \tilde{\beta}^{\lambda}_{-1} +
\nonumber\\
{1\over 2}\eta^{(6)}_{\mu\nu\lambda} \beta^{\mu}_{-2}  \tilde{\alpha}^{\nu}_{-1}  \tilde{\alpha}^{\lambda}_{-1} +
~
{1\over 2}\eta^{(8)}_{\mu\nu\lambda}   \beta^{\mu}_{-2}  \tilde{\alpha}^{\nu}_{-1}  \tilde{\beta}^{\lambda}_{-1} +
{1\over 2}\eta^{(8)}_{\mu\nu\lambda}   \beta^{\mu}_{-2}  \tilde{\beta}^{\nu}_{-1}  \tilde{\alpha}^{\lambda}_{-1} +
\nonumber\\
{1\over 2}\eta^{(10)}_{\mu\nu\lambda} \alpha^{\mu}_{-1}  \alpha^{\nu}_{-1} \tilde{\alpha}^{\lambda}_{-2} +
~
{1\over 2}\eta^{(12)}_{\mu\nu\lambda}   \alpha^{\mu}_{-1} \beta^{\nu}_{-1}    \tilde{\alpha}^{\lambda}_{-2} +
{1\over 2}\eta^{(12)}_{\mu\nu\lambda}   \beta^{\mu}_{-1} \alpha^{\nu}_{-1}    \tilde{\alpha}^{\lambda}_{-2} +
\nonumber\\
{1\over 4}\eta^{(13)}_{\mu\nu\lambda} \alpha^{\mu}_{-1}  \alpha^{\nu}_{-1}  \tilde{\alpha}^{\lambda}_{-2}+
{1\over 4}\eta^{(14)}_{\mu\nu\lambda} \alpha^{\mu}_{-1}  \beta^{\nu}_{-1} \tilde{\alpha}^{\lambda}_{-2} +
~
{1\over 4}\eta^{(16)}_{\mu\nu\lambda}   \beta^{\mu}_{-1} \beta^{\nu}_{-1}    \tilde{\alpha}^{\lambda}_{-2} +
\nonumber\\
{1\over 2}\eta^{(14)}_{\mu\nu\lambda} \alpha^{\mu}_{-1}  \alpha^{\nu}_{-1} \tilde{\beta}^{\lambda}_{-2} +
~
{1\over 2}\eta^{(16)}_{\mu\nu\lambda}   \alpha^{\mu}_{-1} \beta^{\nu}_{-1}    \tilde{\beta}^{\lambda}_{-2} +
{1\over 2}\eta^{(16)}_{\mu\nu\lambda}   \beta^{\mu}_{-1} \alpha^{\nu}_{-1}    \tilde{\beta}^{\lambda}_{-2} +\nonumber\\
\nonumber\eeqa
plus the part which is quartic in the operators with overall coefficient $1/2$
\beqa
\xi^{(2)}_{\mu\nu\lambda\rho} \alpha^{\mu}_{-1}  \alpha^{\nu}_{-1}\tilde{\alpha}^{\lambda}_{-1}\tilde{\alpha}^{\rho}_{-1} +
~
\xi^{(4)}_{\mu\nu\lambda\rho}  \alpha^{\mu}_{-1}  \alpha^{\nu}_{-1} \tilde{\alpha}^{\lambda}_{-1}\tilde{\beta}^{\rho}_{-1} +
\xi^{(4)}_{\mu\nu\lambda\rho}  \alpha^{\mu}_{-1}  \alpha^{\nu}_{-1} \tilde{\beta}^{\lambda}_{-1}\tilde{\alpha}^{\rho}_{-1} +
\nonumber\\
\xi^{(5)}_{\mu\nu\lambda\rho} \alpha^{\mu}_{-1}  \alpha^{\nu}_{-1}\tilde{\alpha}^{\lambda}_{-1}\tilde{\alpha}^{\rho}_{-1}+
\xi^{(6)}_{\mu\nu\lambda\rho} \alpha^{\mu}_{-1}  \alpha^{\nu}_{-1}\tilde{\alpha}^{\lambda}_{-1}\tilde{\beta}^{\rho}_{-1} +
\xi^{(6)}_{\mu\nu\lambda\rho} \alpha^{\mu}_{-1}  \beta^{\nu}_{-1}\tilde{\alpha}^{\lambda}_{-1}\tilde{\alpha}^{\rho}_{-1} +
~
\nonumber\\
~
\xi^{(8)}_{\mu\nu\lambda\rho}  \alpha^{\mu}_{-1}  \alpha^{\nu}_{-1} \tilde{\beta}^{\lambda}_{-1}\tilde{\beta}^{\rho}_{-1} +
\xi^{(8)}_{\mu\nu\lambda\rho}  \alpha^{\mu}_{-1}  \beta^{\nu}_{-1} \tilde{\alpha}^{\lambda}_{-1}\tilde{\beta}^{\rho}_{-1} +
\xi^{(8)}_{\mu\nu\lambda\rho}  \alpha^{\mu}_{-1}  \beta^{\nu}_{-1} \tilde{\beta}^{\lambda}_{-1}\tilde{\alpha}^{\rho}_{-1} +
\nonumber\\
\xi^{(13)}_{\mu\nu\lambda\rho} \alpha^{\mu}_{-1}  \beta^{\nu}_{-1}\tilde{\alpha}^{\lambda}_{-1}\tilde{\alpha}^{\rho}_{-1}+
\xi^{(13)}_{\mu\nu\lambda\rho} \beta^{\mu}_{-1}  \alpha^{\nu}_{-1}\tilde{\alpha}^{\lambda}_{-1}\tilde{\alpha}^{\rho}_{-1}+
\nonumber\\
\xi^{(14)}_{\mu\nu\lambda\rho} \alpha^{\mu}_{-1}  \beta^{\nu}_{-1}\tilde{\alpha}^{\lambda}_{-1}\tilde{\beta}^{\rho}_{-1} +
\xi^{(14)}_{\mu\nu\lambda\rho} \beta^{\mu}_{-1}  \alpha^{\nu}_{-1}\tilde{\alpha}^{\lambda}_{-1}\tilde{\beta}^{\rho}_{-1} +
\xi^{(14)}_{\mu\nu\lambda\rho} \beta^{\mu}_{-1}  \beta^{\nu}_{-1}\tilde{\alpha}^{\lambda}_{-1}\tilde{\alpha}^{\rho}_{-1} +
~
\nonumber\\
\xi^{(16)}_{\mu\nu\lambda\rho}  \alpha^{\mu}_{-1}  \beta^{\nu}_{-1} \tilde{\beta}^{\lambda}_{-1}\tilde{\beta}^{\rho}_{-1} +
\xi^{(16)}_{\mu\nu\lambda\rho}  \beta^{\mu}_{-1}  \alpha^{\nu}_{-1} \tilde{\beta}^{\lambda}_{-1}\tilde{\beta}^{\rho}_{-1} +
\xi^{(16)}_{\mu\nu\lambda\rho}  \beta^{\mu}_{-1}  \beta^{\nu}_{-1} \tilde{\alpha}^{\lambda}_{-1}\tilde{\beta}^{\rho}_{-1} +
\xi^{(16)}_{\mu\nu\lambda\rho}  \beta^{\mu}_{-1}  \beta^{\nu}_{-1} \tilde{\beta}^{\lambda}_{-1}\tilde{\alpha}^{\rho}_{-1}~~~
\nonumber\\
|k,e,0> =0.~~~~~~\nonumber
\eeqa
The sum of the coefficients in front of the identical operators should be equal to zero.
Taking into account the symmetries of individual tensors in $|\Psi_2>$ one can get the
following equations:
\beqa
\begin{array}{l}
\chi^{(4)} =0,~~~~~~~~~~~~~~~~~~\chi^{(2)}  +\chi^{(3)} =0,\nonumber\\\\
\eta^{(8)} =0,~~~~~~~~~~~~~~~~~~~\eta^{(6)}=0,~~~~~~~~~~~~~{1\over 2}\eta^{(2)} +
{1\over 4}\eta^{(5)}=0,\nonumber\\
\eta^{(4)} ~+~ {1\over 4}\eta^{(6)} ~=~0,~~~~~\nonumber\\
\eta^{(16)}=0,~~~~~~~~~~~~~~~~~~\eta^{(14)}=0,~~~~~~~~~~~{1\over 2}\eta^{(10)} +
{1\over 4}\eta^{(13)}=0,\nonumber\\
\eta^{(12)}  ~+~ {1\over 4}\eta^{(14)}=0,
\nonumber\\\\
\xi^{(2)} +\xi^{(5)}  =0,~~~~~\nonumber\\
2 \xi^{(4)} +\xi^{(6)} =0,~~~~~~~~~~~~~~~~~~~~
\xi^{(6)} +2\xi^{(13)} =0,
\nonumber\\
\xi^{(8)} =0,~~~~~~~~~~~~~~~~~~~~~~~~~~~~~~\xi^{(14)} =0,
\nonumber\\
2\xi^{(8)} +\xi^{(14)} =0~~~~~~~~~~~~~~~~~~~
~~\xi^{(16)} =0.
\end{array}
\eeqa
This system of linear equations can be easily solved. As one can see,
those tensors which are connected with the structures having
three and four $\beta$ operators, in the product, are equal to zero.
This phenomenon already appeared when we analyzed the first level wave function:
the tensor which is in front of the two $\beta$ operators was
equal to zero\cite{Savvidy:2003fx}.
The solution can be expressed in the form
\beqa
\eta^{(4)} = \eta^{(6)}  =\eta^{(8)} =
\eta^{(12)} = \eta^{(14)} =\eta^{(16)} =0,~~~~~~~~~~~~~~~~~~~~~~~~~~\nonumber\\
\eta^{(5)} = -2 \eta^{(2)} ,~~~~~~~~~~~
\eta^{(13)} = -2 \eta^{(10)} ,~~~~~~~~~~~~~~~~~~~~~~~~~~~~\nonumber\\
\xi^{(8)}=\xi^{(14)} = \xi^{(16)}=0,~~~\xi^{(5)}= -\xi^{(2)},~~~
\xi^{(6)}=-2\xi^{(4)}=-2\xi^{(13)}.\nonumber
\eeqa
Therefore, after imposing the $\Theta_{00}$ constraint, the wave function
has three sets of terms:
\beqa
|\Psi_2> ~~~~~~~~~~~~~~~~~~= ~~~~~~~~~~~~~
\chi^{(1)}_{\mu\nu} ~\alpha^{\mu}_{-2}  \tilde{\alpha}^{\nu}_{-2}
+ \chi^{(2)}_{\mu\nu}  ~(\alpha^{\mu}_{-2} \tilde{\beta}^{\nu}_{-2} -
\beta^{\mu}_{-2}  \tilde{\alpha}^{\nu}_{-2})+
\nonumber\\\nonumber\\
~\eta^{(1)}_{\mu\nu\lambda} ~\alpha^{\mu}_{-2}  \tilde{\alpha}^{\nu}_{-1} \tilde{\alpha}^{\lambda}_{-1}+
\eta^{(2)}_{\mu\nu\lambda}~ \alpha^{\mu}_{-2}  \tilde{\alpha}^{\nu}_{-1}  \tilde{\beta}^{\lambda}_{-1} +
~
~\eta^{(5)}_{\mu\nu\lambda} ~\beta^{\mu}_{-2}  \tilde{\alpha}^{\nu}_{-1} \tilde{\alpha}^{\lambda}_{-1}+
\nonumber\\
~\eta^{(9)}_{\mu\nu\lambda} ~\alpha^{\mu}_{-1}  \alpha^{\nu}_{-1}  \tilde{\alpha}^{\lambda}_{-2}
+ \eta^{(10)}_{\mu\nu\lambda}~ \alpha^{\mu}_{-1}  \beta^{\nu}_{-1} \tilde{\alpha}^{\lambda}_{-2} +
~
\eta^{(13)}_{\mu\nu\lambda} ~\alpha^{\mu}_{-1}  \alpha^{\nu}_{-1}  \tilde{\beta}^{\lambda}_{-2}+
\nonumber\\
\xi^{(1)}_{\mu\nu\lambda\rho} \alpha^{\mu}_{-1}  \alpha^{\nu}_{-1}\tilde{\alpha}^{\lambda}_{-1}\tilde{\alpha}^{\rho}_{-1}
+ \xi^{(2)}_{\mu\nu\lambda\rho} ~\alpha^{\mu}_{-1}  \alpha^{\nu}_{-1}\tilde{\alpha}^{\lambda}_{-1}\tilde{\beta}^{\rho}_{-1} +
\xi^{(5)}_{\mu\nu\lambda\rho} ~\alpha^{\mu}_{-1}  \beta^{\nu}_{-1}\tilde{\alpha}^{\lambda}_{-1}\tilde{\alpha}^{\rho}_{-1} +
\nonumber\\\nonumber\\
\xi^{(4)}_{\mu\nu\lambda\rho} ~( \alpha^{\mu}_{-1} \beta^{\nu}_{-1}\tilde{\alpha}^{\lambda}_{-1}\tilde{\beta}^{\rho}_{-1} +
\beta^{\mu}_{-1}  \alpha^{\nu}_{-1} \tilde{\beta}^{\lambda}_{-1}\tilde{\alpha}^{\rho}_{-1} -
\alpha^{\mu}_{-1}  \alpha^{\nu}_{-1} \tilde{\beta}^{\lambda}_{-1}\tilde{\beta}^{\rho}_{-1}-
\beta^{\mu}_{-1} \beta^{\nu}_{-1}\tilde{\alpha}^{\lambda}_{-1}\tilde{\alpha}^{\rho}_{-1} )+
|k,e,0>, \nonumber
\eeqa
where still we have the conditions $\eta^{(5)} = -2 \eta^{(2)} ,~
\eta^{(13)} = -2 \eta^{(10)}$ and
$\xi^{(2)} + \xi^{(5)} =0$.
Resolving these last equations we get nine independent tensors:
\beqa\label{secondlevelwave}
|\Psi_2> ~~~~~~~~~~~~~~~~~~= ~~~~~~~~~~~~~
\chi^{(1)}_{\mu\nu} ~\alpha^{\mu}_{-2}  \tilde{\alpha}^{\nu}_{-2}
+ \chi_{\mu\nu}  ~(\alpha^{\mu}_{-2} \tilde{\beta}^{\nu}_{-2} -
\beta^{\mu}_{-2}  \tilde{\alpha}^{\nu}_{-2})+
\nonumber\\\\
~\eta^{(1)}_{\mu\nu\lambda} ~\alpha^{\mu}_{-2}  \tilde{\alpha}^{\nu}_{-1} \tilde{\alpha}^{\lambda}_{-1}+
\eta_{\mu\nu\lambda}~ (\alpha^{\mu}_{-2}  \tilde{\alpha}^{\nu}_{-1}  \tilde{\beta}^{\lambda}_{-1} -
2~\beta^{\mu}_{-2}  \tilde{\alpha}^{\nu}_{-1} \tilde{\alpha}^{\lambda}_{-1})+
\nonumber\\
~\zeta^{(1)}_{\mu\nu\lambda} ~\alpha^{\mu}_{-1}  \alpha^{\nu}_{-1}  \tilde{\alpha}^{\lambda}_{-2}
+ \zeta_{\mu\nu\lambda}~(\beta^{\mu}_{-1} \alpha^{\nu}_{-1} \tilde{\alpha}^{\lambda}_{-2} -
2~\alpha^{\mu}_{-1}  \alpha^{\nu}_{-1}  \tilde{\beta}^{\lambda}_{-2})+
\nonumber\\\nonumber\\
\xi^{(1)}_{\mu\nu\lambda\rho} \alpha^{\mu}_{-1}  \alpha^{\nu}_{-1}\tilde{\alpha}^{\lambda}_{-1}\tilde{\alpha}^{\rho}_{-1}
+ \xi^{(2)}_{\mu\nu\lambda\rho} ~(\alpha^{\mu}_{-1} \alpha^{\nu}_{-1}\tilde{\alpha}^{\lambda}_{-1}\tilde{\beta}^{\rho}_{-1}
-~\beta^{\mu}_{-1} \alpha^{\nu}_{-1} \tilde{\alpha}^{\lambda}_{-1}\tilde{\alpha}^{\rho}_{-1} )+
\nonumber\\
\xi_{\mu\nu\lambda\rho} ~( \alpha^{\mu}_{-1} \beta^{\nu}_{-1}\tilde{\alpha}^{\lambda}_{-1}\tilde{\beta}^{\rho}_{-1} +
\beta^{\mu}_{-1}  \alpha^{\nu}_{-1} \tilde{\beta}^{\lambda}_{-1}\tilde{\alpha}^{\rho}_{-1} -
\alpha^{\mu}_{-1}  \alpha^{\nu}_{-1} \tilde{\beta}^{\lambda}_{-1}\tilde{\beta}^{\rho}_{-1}-
\beta^{\mu}_{-1} \beta^{\nu}_{-1}\tilde{\alpha}^{\lambda}_{-1}\tilde{\alpha}^{\rho}_{-1} )
|k,e,0>,~~~\nonumber
\eeqa
where it was convenient to introduce the tensors
$$\xi^{(4)}_{\mu\nu\lambda\rho} =\xi_{\mu\nu\lambda\rho},~~~~~~
\eta^{(2)}_{\mu\nu\lambda}= \eta_{\mu\nu\lambda},~~~~~~
\eta^{(10)}_{\mu\nu\lambda} = \zeta_{\mu\nu\lambda},~~~~
\chi^{(2)}_{\mu\nu} =\chi_{\mu\nu}
$$
in order to sum up the similar terms. The tensor
$\xi_{\mu\nu\lambda\rho}$ is symmetric under simultaneous interchange
of the indices $\mu \leftrightarrow \nu$ and $\lambda \leftrightarrow \rho$.
One can simply check a posteriori that the wave function (\ref{secondlevelwave})
fulfills the constrain $\Theta_{00}|\Psi_1>=0$.

Our aim is now to impose
the rest of the constraints on the wave function (\ref{secondlevelwave}).
The next constraint to be imposed is $\Theta_{10}|\Psi_2>=0$:
\beqa
\Theta_{10}|\Psi_2>= -{1\over 4}\chi_{\mu\nu}  \alpha^{\mu}_{-1}
\tilde{\alpha}^{\nu}_{-2} - {1\over 2} \eta_{\mu\nu\lambda} \alpha^{\mu}_{-1}
\tilde{\alpha}^{\nu}_{-1} \tilde{\alpha}^{\lambda}_{-1}+ i(\zeta_{\mu\nu\lambda}
e^{\mu} \alpha^{\nu}_{-1}  \tilde{\alpha}^{\lambda}_{-2} -
\xi^{(2)}_{\mu\nu\lambda\rho}
e^{\mu} \alpha^{\nu}_{-1}  \tilde{\alpha}^{\lambda}_{-1}\tilde{\alpha}^{\rho}_{-1}
+ \nonumber\\
\xi_{\mu\nu\lambda\rho}( \alpha^{\mu}_{-1} e^{\nu}
\tilde{\alpha}^{\lambda}_{-1}\tilde{\beta}^{\rho}_{-1} +
e^{\mu}\alpha^{\nu}_{-1}
\tilde{\beta}^{\lambda}_{-1} \tilde{\alpha}^{\rho}_{-1}-
e^{\mu}\beta^{\nu}_{-1}\tilde{\alpha}^{\lambda}_{-1}\tilde{\alpha}^{\rho}_{-1}-
\beta^{\mu}_{-1}e^{\nu}\tilde{\alpha}^{\lambda}_{-1}\tilde{\alpha}^{\rho}_{-1})
)|k,e,0> =0~~~\nonumber
\eeqa
and similar expressions that follow for the constraints
$\Theta_{01}|\Psi_2>=\Theta_{11}|\Psi_2> = 0$. All equations which follow
from these constraints can be summarized  in the following form
\beqa\label{theta1}
e^{\mu}~ \xi_{\mu\nu\lambda\rho} =e^{\lambda}~ \xi_{\mu\nu\lambda\rho} =0,~~~~~~~~~~~~~~\nonumber\\
\xi_{\mu\lambda\nu\lambda} =0,~~~~~~~~~~~~~~~~~~~~~~~~~~\nonumber\\
i  ~e^{\mu}~\zeta_{\mu\nu\lambda}
-{1\over 4} \chi_{\nu\lambda}=0,~~~
i  ~e^{\mu}~\xi^{(2)}_{\mu\nu\lambda\rho}+
{1\over 2} \eta_{\nu\lambda\rho}=0,\nonumber\\
i ~e^{\lambda}~\eta_{\mu\nu\lambda}
+{1\over 4} \chi_{\mu\nu}=0,~~~
i ~e^{\rho}~\xi^{(2)}_{\mu\nu\lambda\rho}-
{1\over 2} \zeta_{\mu\nu\lambda}=0.
\eeqa
Imposing the equations $\Theta_{20}|\Psi_2>=
\Theta_{02}|\Psi_2>=\Theta_{22}|\Psi_2> = 0$ we get
\beqa\label{theta2}
e^{\mu} \chi_{\mu\nu}=\chi_{\mu\nu}e^{\nu} =0,~~~~~~~~~~~~~~~~~~~\nonumber\\
i  e^{\mu}\eta_{\mu\lambda\rho} -{1\over 4} \xi_{\mu\mu\lambda\rho}=0,~~~
i e^{\lambda}\zeta_{\mu\nu\lambda} -{1\over 4} \xi_{\mu\nu\lambda\lambda}=0,
\eeqa
while $\Theta_{12}|\Psi_2>=\Theta_{21}|\Psi_2>= 0$ does not produce
any new equations. Finally, imposing the constraints $k\cdot \alpha_1 =
k\cdot \tilde{\alpha_1} =  k\cdot \alpha_2 = k\cdot \tilde{\alpha_2} =0$ we
get
\beqa\label{theta3}
k^{\mu} ~\xi_{\mu\nu\lambda\rho}  &=&
\xi_{\nu\mu\rho\lambda} k^{\rho} =0,~~~~~~~~~~~~~\nonumber\\
k^{\mu} \chi_{\mu\nu}~&=&~\chi_{\mu\nu}k^{\nu} =0,~~~~~~~~~~~~~~~\nonumber\\
k^{\lambda}\eta_{\lambda\mu\nu} &=& \eta_{\mu\nu\lambda}k^{\lambda}=0,
~~~~~~~~~~~~~~~\nonumber\\
k^{\lambda}\zeta_{\lambda\mu\nu}&=&
\zeta_{\mu\nu\lambda}k^{\lambda}=0,~~~~~~~~~~~~~~~\nonumber\\
k^{\lambda}\xi^{(2)}_{\lambda\mu\nu\rho}&=&
\xi^{(2)}_{\mu\nu\rho\lambda}k^{\lambda}=0.
\eeqa
Thus we have imposed all constraints associated with the $\Theta$
operator, corresponding to the equations
(\ref{theta1}),(\ref{theta2}) and (\ref{theta3}).

Now we are in a position to consider the Virasoro constraints. The
$L_2 |\Psi_2> = \tilde{L}_2 |\Psi_2> =0$ equations imply that
\beqa\label{l2equations}
k^\mu \chi^{(1)}_{\mu\nu} - \pi^{\mu}\chi_{\mu\nu}~=0,~~~~~~~~~~~
k^\nu \chi^{(1)}_{\mu\nu} + \pi^{\nu}\chi_{\mu\nu} ~=0,\nonumber\\
k^\mu \eta^{(1)}_{\mu\nu\lambda} - 2\pi^{\mu} \eta_{\mu\nu\lambda}~=0,~~~~~~~~
k^\lambda \zeta^{(1)}_{\mu\nu\lambda} - 2\pi^{\lambda}\zeta_{\mu\nu\lambda} ~=0,\nonumber\\
\xi_{\lambda\lambda\mu\nu}=\xi_{\mu\nu\lambda\lambda} =0.~~~~~~~~~~~~~~~~~~~~
\eeqa
Finally we came to the last constraints, $L_1 |\Psi_2> = 0$
which imply that
\beqa\label{l1equation}
k^\mu \zeta_{\nu\mu\lambda} =0,~~~~~~~~~~~~~~~~~~~~~~~~\chi_{\nu\lambda} ~=0\nonumber\\
\eta_{\nu\lambda\rho}~=0,~~~~~~~~~
k^\mu \xi^{(2)}_{\nu\mu\lambda\rho} + \pi^{\mu}\xi_{\mu\nu\lambda\rho} =0 \nonumber\\
k^\mu (\zeta^{(1)}_{\mu\nu\lambda} ~+ ~\zeta^{(1)}_{\nu\mu\lambda} )  +
\pi^{\mu} \zeta_{\mu\nu\lambda}~ +~ 2\chi^{(1)}_{\nu\lambda}=0 \nonumber\\
k^\mu (\xi^{(1)}_{\mu\nu\lambda\rho} + \xi^{(1)}_{\nu\mu\lambda\rho} )  -
\pi^{\mu} \xi^{(2)}_{\mu\nu\lambda\rho}  + 2 \eta^{(1)}_{\nu\lambda\rho}  =0
\eeqa
and $\tilde{L}_1 |\Psi_2> =0$, that gives:
\beqa\label{l1tildaequation}
k^\mu \eta_{\nu\mu\lambda}=0,~~~~~~~~~~~~~~~~~~~~~~~\chi_{\nu\lambda} ~=0 \nonumber\\
\zeta_{\nu\lambda\rho}~=0,~~~~~~~~~~
k^\mu \xi^{(2)}_{\nu\lambda\mu\rho} - \pi^{\mu}\xi_{\nu\lambda\mu\rho} =0 \nonumber\\
k^\mu (\eta^{(1)}_{\nu\mu\lambda} ~+~ \eta^{(1)}_{\nu\lambda\mu} ) ~ +~
\pi^{\mu} \eta_{\nu\lambda\mu} + 2\chi^{(1)}_{\nu\lambda}=0 \nonumber\\
k^\mu (\xi^{(1)}_{\nu\lambda\rho\mu} + \xi^{(1)}_{\nu\lambda\mu\rho} )  +
\pi^{\mu} \xi^{(2)}_{\nu\lambda\rho\mu}  + 2 \zeta^{(1)}_{\nu\lambda\rho}  =0.
\eeqa

\section{{\it Solution of the Constraints}}

Our system of equations (\ref{theta1}),(\ref{theta2}), (\ref{theta3})
and (\ref{l2equations}),(\ref{l1equation}),(\ref{l1tildaequation}),
which define the wave function (\ref{secondlevelwave}), further reduces
to the form:
\beqa\label{secondlevelwave1}
|\Psi_2> ~~~~~~~~~~~~~~~~~~= ~~~~~~~~~~~~~
\chi^{(1)}_{\mu\nu} ~\alpha^{\mu}_{-2}  \tilde{\alpha}^{\nu}_{-2} +
\eta^{(1)}_{\mu\nu\lambda} ~\alpha^{\mu}_{-2}\tilde{\alpha}^{\nu}_{-1}\tilde{\alpha}^{\lambda}_{-1}+
+\zeta^{(1)}_{\mu\nu\lambda} ~\alpha^{\mu}_{-1}  \alpha^{\nu}_{-1}  \tilde{\alpha}^{\lambda}_{-2}
\nonumber\\
\xi^{(1)}_{\mu\nu\lambda\rho} \alpha^{\mu}_{-1}  \alpha^{\nu}_{-1}\tilde{\alpha}^{\lambda}_{-1}\tilde{\alpha}^{\rho}_{-1}
+ \xi^{(2)}_{\mu\nu\lambda\rho} ~(\alpha^{\mu}_{-1} \alpha^{\nu}_{-1}\tilde{\alpha}^{\lambda}_{-1}\tilde{\beta}^{\rho}_{-1}
-~\beta^{\mu}_{-1} \alpha^{\nu}_{-1} \tilde{\alpha}^{\lambda}_{-1}\tilde{\alpha}^{\rho}_{-1} )+
\nonumber\\
\xi_{\mu\nu\lambda\rho} ~( \alpha^{\mu}_{-1} \beta^{\nu}_{-1}\tilde{\alpha}^{\lambda}_{-1}\tilde{\beta}^{\rho}_{-1} +
\beta^{\mu}_{-1}  \alpha^{\nu}_{-1} \tilde{\beta}^{\lambda}_{-1}\tilde{\alpha}^{\rho}_{-1} -
\alpha^{\mu}_{-1}  \alpha^{\nu}_{-1} \tilde{\beta}^{\lambda}_{-1}\tilde{\beta}^{\rho}_{-1}-
\beta^{\mu}_{-1} \beta^{\nu}_{-1}\tilde{\alpha}^{\lambda}_{-1}\tilde{\alpha}^{\rho}_{-1} )
|k,e,0>,
\eeqa
where we have used equations (\ref{l1equation}) and (\ref{l1tildaequation}),
$\chi_{\mu\nu}=~ \zeta_{\mu\nu\lambda}~=\eta_{\mu\nu\lambda}~=0$.
We shall summarize all remaining constraints in the form:
\beqa\label{theta11}
e^{\mu}~ \xi_{\mu\nu\lambda\rho} =e^{\lambda}~ \xi_{\mu\nu\lambda\rho} =0,~~~~~~~~~~~~~~~~~~~~~~~~~~~~~~~~~~~~~~~~~~~~~~~
k^{\mu} ~\xi_{\mu\nu\lambda\rho}  =
\xi_{\nu\mu\rho\lambda} k^{\rho} =0,&\nonumber\\
\xi_{\mu\lambda\nu\lambda} =0,~~~~ \xi_{\lambda\lambda\mu\nu}=0,~~~~~
\xi_{\mu\nu\lambda\lambda}=0,~~~~~~~~~~~~~~~~~~~~~~~~~~~~&\nonumber\\
e^{\mu}~\xi^{(2)}_{\mu\nu\lambda\rho} =e^{\rho}~\xi^{(2)}_{\mu\nu\lambda\rho} =0,~~~~~~~~~~~~~~~~~~~~~~~~~~~~~~~~~~~~~~~~~~~~~~~~
k^{\lambda}~\xi^{(2)}_{\lambda\mu\nu\rho}=
\xi^{(2)}_{\mu\nu\rho\lambda}k^{\lambda}=0,&\nonumber\\
k^\mu \xi^{(2)}_{\mu\nu\lambda\rho} + \pi^{\mu}\xi_{\mu\nu\lambda\rho} =0,~~~~~~~~~~~~~~~~~~~~~~~~~~~~~~~~~~~~~~~~~~~~~~~~~~
k^\mu ~\xi^{(2)}_{\nu\lambda\mu\rho} - \pi^{\mu}\xi_{\nu\lambda\mu\rho} =0& \nonumber\\
2k^\mu \xi^{(1)}_{\mu\nu\lambda\rho}  -
\pi^{\mu} \xi^{(2)}_{\mu\nu\lambda\rho}  + 2 \eta^{(1)}_{\nu\lambda\rho}  =0,~~~~~~~~~~~~~~~~~~~~~~~~~~~~
2k^\mu \xi^{(1)}_{\nu\lambda\rho\mu}  +
\pi^{\mu} \xi^{(2)}_{\nu\lambda\rho\mu}  + 2 \zeta^{(1)}_{\nu\lambda\rho}  =0&\nonumber\\
k^\mu \eta^{(1)}_{\mu\nu\lambda}~=0,~~~~~~~~~~~~~~~~~~~~~~~~~~~~~~~~~~~~~~~~~~~~~~~~~~~~~~~~~~~~~~~~~~~~~~~~~~~~~~
k^\lambda \zeta^{(1)}_{\mu\nu\lambda} ~=0&\nonumber\\
k^\mu \eta^{(1)}_{\nu\lambda\mu}   +  \chi^{(1)}_{\nu\lambda}=0,~~~~~~~~~~~~~~~~~~~~~~~~~~~~~~~~~~~~~~~~~~~~~~~~~~~~~~~~~~~~~~
k^\mu \zeta^{(1)}_{\mu\nu\lambda}  +
\chi^{(1)}_{\nu\lambda}=0&\nonumber\\
k^\mu \chi^{(1)}_{\mu\nu}~=k^\nu \chi^{(1)}_{\mu\nu} ~=0~~~~~~~~~~~~~~~~~~~~~~~~~~~~~~~~~~~~~~~~~&
\eeqa
It is appropriate now to compute the norm of the wave function $|\Psi_2> $:
\be
<\Psi_2 |\Psi_2> = 12\xi^{*}_{\mu\nu\lambda\rho}\xi_{\mu\nu\lambda\rho}.
\ee
As one can see only one tensor $\xi_{\mu\nu\lambda\rho}(k,e)$
contributes into the norm of the second level wave function.
The rest of the tensors $\chi^{(1)}_{\mu\nu}$,
$\eta^{(1)}_{\mu\nu\lambda},~\zeta^{(1)}_{\mu\nu\lambda}$ and
$\xi^{(1,2)}_{\mu\nu\lambda\rho}$ do not contribute to the norm.
We have the following equations on this basic tensor $\xi_{\mu\nu\lambda\rho}(k,e)
= \xi_{\nu\mu\rho\lambda}(k,e)$
\beqa
k^{\mu} \xi_{\mu\nu\lambda\rho}=\xi_{\mu\nu\lambda\rho}k^{\lambda}
=0,~~~e^{\mu} \xi_{\mu\nu\lambda\rho}=\xi_{\mu\nu\lambda\rho}e^{\lambda} =0,~~~\nonumber\\
\xi_{\mu\lambda\nu\lambda} =0,~~~~ \xi_{\lambda\lambda\mu\nu}=0,~~~~~
\xi_{\mu\nu\lambda\lambda}=0,~~~~~~~~~
\eeqa
which generalize the corresponding equations on $\omega_{\mu\nu}$
for the first excited state $\Xi=1$ where ~$k^\mu \omega_{\mu\nu} =
\omega_{\mu\nu} k^\nu =0,~
e^\mu \omega_{\mu\nu} = \omega_{\mu\nu} e^\nu =0$.
They tell us again that the tensor $\xi$ 
should be transverse and longitudinal at the same time.
Unique solution of these equations therefore is:
\be
\xi_{\mu\nu\lambda\rho} =  \xi(k,e)~ k^{\mu} k^{\nu} k^{\lambda} k^{\rho},
\ee
from which it follows that the norm of the second level wave function
is equal to zero
\be
<\Psi_2 |\Psi_2> = 0.
\ee

One of the authors (G.S.) wishes to thank CERN Theory Division for
hospitality and Luis Alvarez-Gaume, Spenta Wadia, Kumar Narain
and Kostas Kounnas for stimulating discussions. This work was
supported in part by the EEC Grant no. HPRN-CT-1999-00161.

\vfill
\end{document}